\documentclass[pre,aps,twocolumn]{revtex4-1}
\usepackage{graphicx,mathtools,morefloats}
\makeatletter
\newcommand*{\centerfloat}{%
  \parindent \z@
  \leftskip \z@ \@plus 1fil \@minus \textwidth
  \rightskip\leftskip
  \parfillskip \z@skip}
\makeatother

\begin{document}

\title{Modelling Statistics of Polypeptides in Emissions from
Smokers Near Ocean Ridges}

\author{Ben Intoy  and J. W. Halley}
\address{School of Physics and
Astronomy, University of Minnesota, Minneapolis, MN 55455}

\date{\today}

\begin{abstract}
We have previously shown in model studies that rapid quenches
of systems of monomers interacting to form polymer chains can fix
nonequilibrium chemistries which could lead to the origin of life.
We suggested that such quenching processes might have occurred
at very high rates on early earth, giving an efficient mechanism
for natural sorting through enormous numbers of nonequilibrium
chemistries from which those most likely to lead to life could
be  naturally selected.
Taking account of kinetic barriers, we found good agreement between
laboratory quenching experiments on solutions of amino acids 
and the resulting model.  We also
made a preliminary comparison between reported data on polypeptides sampled
from emissions from smokers near ocean ridges and our model. 
However that previous model
assumed that the concentrations of all monomeric amino acids in the medium
were the same whereas that is not the case in  samples taken from 
ocean smokers. Here we take account of the heterogeneous concentrations
of the amino acid monomers in the medium in the analysis of the smoker data and compare the results with the data on polypeptide concentrations found
in the samples taken by a submersible in the Marianna Trough. Results
are consistent with the hypothesis that smokers were the 
source of large and extremely diverse number of polypeptides in thermal
disequilibrium which could incubate processes leading to life.  
\end{abstract}
 
\pacs{PACS numbers:  }

\maketitle

\section{Introduction} \label{sec:Introduction}

The likelihood of natural formation of an initial genome in
'genome first' models of prebiotic evolution
appears to be nearly impossible \cite{eigen}.
That has  motivated interest in alternative models
in which the early phases of prebiotic systems were characterized by
collections of polymers exhibiting lifelike behavior and storing
information collectively without a central  genome.
Estimates
of the likelihood of the random natural formation of
such entities, of which  prions and  amyloids \cite{mauryprions},\cite{reisnerprions}.\cite{baskakovprions},\cite{portilloprions},\cite{greenwald}are often mentioned
as examples, are probably higher but detailed understanding of
how likely they are to form prebiotically is poorly understood and a major issue
in evaluating such models.

Here we address the necessary first step, namely the identification 
of a source of macromolecular polypeptides far from equilibrium which 
could incubate the formation of such prebiotic systems.  
In previous papers\cite{wynveen},\cite{conditions},\cite{quench},\cite{life},\cite{life2},\cite{intoy} we showed that
coarse grained models of putative polymer prebiotic chemistry suggested
that quenching of a collection of interacting amino acids  from a high
temperature to ambient conditions could  allow a  wide natural
exploration of the space of polymer combinations in the high temperature phase
followed by fixing of many diverse  nonequilibrium states some of which could 
have led to the initiation of life-like entities, possibly like amyloids or
prions. Such quenches  occur continuously in the fluids emitted continuously
from fluids emitted from 'smokers' found on ocean ridges on ocean floors.
Those fluids have been proposed as possible locations of terrestrial life's
origin for other reasons for a long time\cite{schreibera,schreiberb,shock}. 
Including a simple account of kinetic
barriers to peptide hydrolysis\cite{life2} in the model of \cite{quench} allowed us to make  a quantitative description of results from laboratory quench 
experiments
\cite{yin},\cite{matsunoa,matsunob,matsunoc}.  We also briefly discussed 
the use of the same model to account for measured concentrations of polypeptides found in 
samples from smokers in the Mariana Trough
\cite{trench}.  However a more
quantitative analysis requires a reformulation of our statistical model
in which the concentrations of the various available monomeric amino acids
are allowed to be  different from one another, whereas our model in reference
\cite{life2} assumed that those
concentrations are all the same. Here we report a modification of the 
model of  reference \cite{life2} which corrects that defect and 
we compare the results
with the observational oceanic data.

We assume the same model for polymer formation which was used in 
\cite{quench,life2}.  In \cite{life2} we argued that the temperatures
and dwell times of the fluid emitted from  the smokers before it was  emitted
were such that the polymer length distributions in the fluid after 
quench in the cold water at the ocean floor would be essentially the
equilibrium distributions that they had before quench. Therefore we will
assume here, as we did in \cite{life2},that the polymer length distribution
of the polypeptides in the fluid after emission into the cold ocean
water remains the same as the thermally equilibrated length distribution
that it had with at a high temperature\cite{seyfriedchapter} just before quench. The thermally 
initiated  breaking of peptide bonds in the hot fluid is 'frozen out'
after quench, leaving the length distribution fixed in a nonequilibrium
state in the cold environment after quench. This assumption is consistent
with our rate estimates based on reported laboratory data and also with 
our analysis reported in \cite{intoyhalleypopulations}
of the polypeptide length distribution
in modern prokaryotes showing that their length distribution is very
nearly the same as that expected at tempertures much higher than the
temperatures at which those organisms currently live. 
With this assumption the analysis of the data from smokers does not
require results from the kinetic models of the dynamics associated
with covalent bond scission and ligation which we used in \cite{life2}
and earlier papers
to analyze laboratory quench experiments because the very high
temperatures of the fluid before quench can be assumed to
have achieved thermal equilibration of the polypeptide lengths
while there can only be negligble peptide bond changes after
quench.  The main difference
between the state before the quench and after it is   
that the numbers of polymers of each length is assumed to be the
same before and after quench but the rate of scission and ligation is
different and much slower in the cold state after quench. Thus the main
analytical problem in analyzing the oceanographic data  is to formulate
a description of the equilibrium distribution of polymer lengths in the 
case that the concentrations 
 of available monomers in the medium in which the chemistry is taking place 
are not all
the same.  

\section{Equilibrium with variable available concentrations} \label{sec:Simulations}

We briefly review our earlier analysis and how it has been modified here:
As in \cite{quench},\cite{wynveen},\cite{conditions},\cite{life2} and elsewhere\cite{kauffman},\cite{others},  we represent biopolymers as strings of monomers labelled
by type with a total number of types $b$. 
In dynamic simulations,  the polymers
 undergo scission and ligation.   However, following the arguments in 
\cite{life2} we assume that in the analysis of the oceanographic data
we can forego dynamic simulation and consider that the equilibrium 
distribution in the hot fluid before quench is retained in the cold
fluid after quench as the fluid passes quickly into the cold water
near the ocean floor.  
As in all our previous reports except \cite{intoy}, we assume 
here that the
system is 'well mixed' and no effects of spatial diffusion are considered.
We assign to  any 'polymer' ( string) of length
$L$  an energy $-(L-1)\Delta $ where $\Delta$ is a
real number which is the bonding energy between two monomers. The
'bonding energies' determine the thermodynamic driving force for
bond formation and are negative in the application to peptide bonds.
They are assumed to be the same for pairs of all types of
monomers and  are to be distinguished from the activation energies for
bond breaking introduced in \cite{life2}. 
Thus all realizations of the  model  have a 
 collection of polymers of lengths $L$
with a number
, $N_L$ of each.  In each polymer there are monomers of types $t=1,..,b$.
The total number of monomers $N_m$ is thus $N_m=\sum_L N_L L$.
In the analysis of the data from the ocean trough experiments of interest
here\cite{trench} 
the global concentration of each type is $p_t$ which is fixed by experiment
so that the total number of available monomers  of type $t$ 
is  $p_tN_m$.
with $\sum_{t=1}^b p_t =1$. The 
main difference between the model used here and the one described
in \cite{life2}  is that in \cite{life2}  we  assumed $p_t=1/b$, that is
that all  monomers are equally available during network formation.
That is definitely not the case in the ocean trough observations
reported in \cite{trench}.  The object of the present paper is to report
an analysis which corrects that defect in the analysis of the  
data reported in \cite{trench}. 

As discussed in the introduction and the next section, it turns out that, to analyse the
results of \cite{trench} in terms of the model described so far here,
we only need to obtain expressions for the expected equilbrium distribution
of the quantities $N_L$, so kinetic simulations of the model are not
required for the analysis we  describe. Therefore we restrict 
the description of the details of the modified model to the 
determination of that equilbrium distribution.    
 
We envision populations of polymers occuring in a bath of available monomers
containing $N_m$ monomers in all 
with fractional global concentrations $p_t$  of monomers of type $t=1,...b$. 
The set of monomers connected by covalent pairwise bonds into polymers
of lengths $L$ from the set $1,...,L_{max}$ so that there are $N_L$
polymers of each length $L$ and $\sum_L N_LL =N_m$. 
We introduce a set of 
variables$\{n_{L,i,t}\}$ 
in which  $n_{L,i,t}$ is the  number of monomers of type $t$ in the 
$i$th polymer among those polymers which are of length $L$.
The index $i$ is introduced because in the present case, the energetic
degeneracies of the polymers of a given length $L$ will not all be the
same.   
There will be therefore be $N_L$ values of $i$ for each length $L$
and $\sum_{L,i}n_{L,i,t}=p_tN_m$, $\sum_{L,i,t}n_{L,i,t}=N_m$
,$\sum_{i,t} n_{L,i,t}=N_LL$,$\sum_{t} n_{L,i,t}
=L$. (In general it will not be true that $\sum_{i} n_{L,i,t}=p_tN_LL$ though
that would be consistent with the other sum rules. If it is true then
we have a special case, discussed more physically below, in which there
has been equilibration of the fraction of types in each individual polymer.) 
The probability of a set $\{n_{L,i,t}\}$ in 
a realization of the model for given $L$, given $i$ and $t=1,...b$ is
the multinomial distribution
\begin{equation}
{\cal{P}}_{L}={{L!\prod_{t=1}^b p_t^{n_{L,i,t}}}\over{\prod_{t=1}^b n_{L,i,t}!}}
\label{eq:PL}
\end{equation}
normalized by summing over all sets $\{n_{L,i,t}\}$ such that $\sum_{t} n_{L,i,t}
=L$.
Previously we did not distinguish the different $i$ values
because they all had the same energetic degeneracy $G_L =b^L$ given 
L.  

 To form a realization of the probability distribution
, we imagine a line  with $N_m$ discrete points.  To assign 
a monomer type to each point, we consider another line of continuous real
values between 0 and 1.  Divide that line into $b$  segments such that the
length of the $t^{th}$ segment is $p_t$.  Now  go through the first line and
, for each of the $N_m$ discrete points,  draw a random number from an even
distribution of random numbers between zero and 1 and assign the discrete point
the type t corresponding to the interval on the unit interval on the 
second line into which the random
number fell. Do this for all the discrete points. The result can be 
regarded as one long polymer of monomer length 
$N_m$. In that long polymer we introduce breaks, corresponding to
broken bonds so that the system is divided into polymers of lengths 
$L$ with $N_L$ polymers of each length.  We assume that, in the high
temperature state before the quench, there is enough time for the
breaks to form (scission) and heal (ligation) until the polymer 
length distribution is equilibrated (that is, has the maximum entropy). 
Notationally, we have
discrete
segments of length $L=1,...,N_m$ such that there are $N_L$ segments of 
length $L$ subject to the overall constraint $\sum_L LN_L =N_m$. 
In such a partitioning, the values of $n_{L,i,t}$ are fixed and
can be determined by counting in the realization and will obey 
the constraints described above. 
We  find the number of ways that that division can take place
and maximize it in order to determine the equilibrium distribution of 
polymer lengths
after quench.
It is the same problem described in our previous work
\cite{quench},\cite{wynveen},\cite{conditions},\cite{life2} 
except that each polymer $L,i$ has a different degeneracy $G_{L,i}$.   
We determine that degeneracy assuming that all arrangements of monomers
within the 
polymer denoted $L,i$  and characterized by the set $\{n_{L,i,t}\}$  
are possible.
Then the number of realizations possible within that set is
\begin{equation}
G_{L,i}={{L!}\over{\prod_{t=1}^b n_{L,i,t}!}}
\label{eq:Gexact}
\end{equation} 

We evaluate the factorials exactly in the numerical calculations
reported here and have, for $L>1$,
$$
G_{L,i}=e^{Ls_{L,i}}
$$
which we can obtain directly from (\ref{eq:Gexact}):
\begin{equation}
s_{L,i}=-\sum_{t}\ln(n_{L,i,t}!)+\ln((L-1)!)
\label{eq:}
\end{equation}
The problem is that the degeneracies are not all the same for 
all the polymers of length $L$ because we have frozen the distribution
of types.  To be sure this is clear consider two subcases where they 
are all the same:  Suppose that each polymer had concentration $p_t$
of type $t$ with $\sum_t p_t=1$.  That physically means we have allowed
the types to rearrange themselves along the line discussed above and that that
process is fast, contrary to what we are going to assume here.    
In that hypothetical case 
$s_{L,i}$ would be independent of $i$ and its value for large $L$ would be 
\begin{equation}
s_L'=-L\sum_{t}p_t\ln p_t
\label{eq:}
\end{equation} 
If we further assumed that all types were equally likely
then $p_t$ would be $ =1/b$ independent of $t$ and it is easy to show that
in that case (which is one we considered in previous papers) we would have
\begin{equation}
s_L''= L\ln b
\label{eq:}
\end{equation} 
and $G_L$ becomes $b^L$ which is the value we have been using
in previous papers. 
To cope with the  present case  in which the $s_{L,i}$ 
can be different for different $i=1,...,N_L$ we note that some of the polymers
 $i$ can have the same value of $s_{L,i}$ for different 
$i$ and will therefore have the same degeneracy.
Some details of the calculation of the polymer length distribution
taking that into account are described in Appendix I.
The result is
\begin{equation}
<\overline{N_L}>={{<e^{sL}>-1}\over {(e^{\beta\Delta(L-1)-\mu\beta}-1)}}
\label{eq:avNLb}
\end{equation} 
for $L\ge 2$ and in which
\begin{equation}
<e^{sL}>=\sum_{\{n_{L,t}\}\ni \sum_t n_{L,t}=L}exp\left[\ln(L-1!)-\sum_t ln (n_{l,t}!)\right]{\cal{ P}}_L
\label{eq:etosLb}
\end{equation} 
where we have dropped the $i$ in $n_{L,i,t}$ because the sum is the same for
all $i$.
$\mu\beta$ is determined  by requiring that $\sum_LL<\overline{N_L}=N_m$. (But $L=1$ must be included separately as explained below. See (\ref{eq:NmN1breakout}).)
If ${{\cal{P}}}_L$  were just a product of delta functions with
  all the $n_{t,L}$ at the values $p_t L$ then this would give 
$e^{L\sum_t-p_t\ln p_t} $ for the degeneracy term as in  the case discussed earlier. For large $L$ $P_L$ has a maximum at that point
but if $L$ is finite, then the average of $e^{sL}$ is less 
 $e^{L\sum_t-p_t\ln p_t} $.  

\section{Analysis of the oceanographic data}
The reported data in \cite{trench} allow evaluation of the ratio of the total
number  of amino acids to the number of monomeric
amino acids  found in each sample.  As discussed in \cite{life2}, the conditions under which the sampling was done as described in \cite{trench} allow us
to use the 'Gibbs limit' in which the -1 in the denominator of 
equation (\ref{eq:avNLb}) is dropped for analysis of the high temperature system before
the quench and the low temperature is low enough so that we can simply
assume that, after quench, the system will retain the equilibrium 
distribution of
polymer lengths that it had in equilibrium at the high temperatures. 
(We are neglecting the possibility of branched polymers here. 
)

\begin{figure}
\includegraphics[width=2.5in,angle=-90]{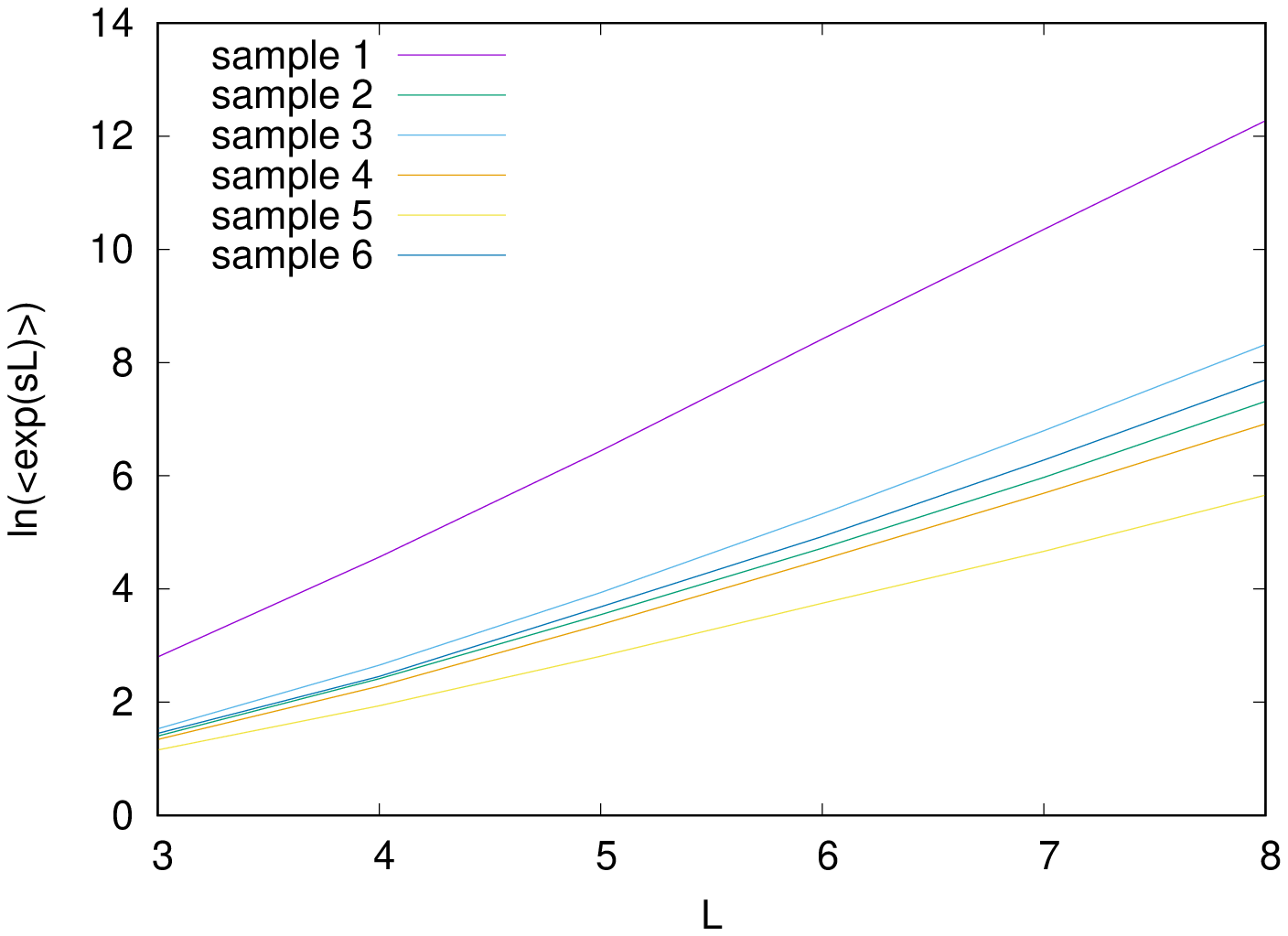}
\caption{\label{fig:lndavvsl}
 .Averaged values $ln(<e^{sL}>)$ for each of the six oceanographic samples
as a function of L.
  }
\end{figure}

Under those conditions the number distribution expected after the quench is
\begin{equation}
<\overline{N_L}>=(<e^{sL}>-1) {(e^{-\beta\Delta(L-1)+\mu\beta})}
\label{eq:avNLgibbs}
\end{equation} 
for $L>1$
where is $\beta$ is evaluated at the high temperature before quench. 
We have evaluated $<e^{sL}>$ numerically using \ref{eq:etosLb} and sampling the 
multinomial distribution using the algoritm described in  
\cite{algo} for 
the distributions $\{p_t,t=1,...,b\}$ empirically observed in the 
samples extracted from the Mariana Trough as reported in 
\cite{trench}. We find that characterisically
\begin{equation}
\ln(<e^{sL}>)\approx C_1(\{p_t)\}L +C_2(\{p_t\}
\label{eq:c1c2}
\end{equation} for $L=3...,8$ as 
shown in Figure \ref{fig:lndavvsl}.

To express the result compactly we define a temperature
$T_{c,2}$ as
\begin{equation}
T_{c,2}\equiv -{{\Delta}\over{k_B C_1(\{p_t\})}}.
\label{eq:}
\end{equation} 
In the case that all the $p_t$ are the same as in the model
studied in \cite{life} , $C_1 $ becomes $\ln b$  
and this definition reduces to 
\begin{equation}
T_{c,2}\equiv -{{\Delta}\over{k_B\ln b}}
\label{eq:oldtc2}
\end{equation}
used in \cite{life2}.  What is different here
is that $T_{c,2}$, which marks the temperature above which the number 
of polymers is predicted to diverge within the model, depends on the 
distribution $\{p_t\}$ and will be different for different smokers for which 
the $\{p_t\}$ are different. As discussed in \cite{life2}, the divergence 
of the prediction for $\sum_L<\overline{N_L}>$ is a consequence of missing 
physics in this and our previous model which must limit the possible maximum 
length of a polymer.  As in \cite{life2} we take account of this by introducing a phenomenological parameter $L_{max}$ which is the maximum possible length
of a polymer due to those unknown physical effects.  However in the analysis
reported here all the observed temperatures of the hot fluids are below 
the values we calculate for $T_{c,2}$ the sums converge  and do not depend
strongly on $L_{max}$ . As in \cite{life2} we 
introduce a parameter  
\begin{equation}
\alpha = \beta\Delta(1-T/T_{c,2})
\label{eq:}
\end{equation} 
 which is formally the same as in \cite{life2}
but with a different definition of $T_{c,2}$
(The overall sign on the right hand side of equation (28)of
\cite{life2} is in error and should be +.). Values of $T_{c,2}$
are different for the six samples studied and are listed in Table 1.

Within the model, there is no energy
associated with totally unbonded monomers so the method of maximization
of the entropy does not provide a prescription for its value.  
We regard the $\mu$ in the previous equations as determined from
the number, termed $N_{m}-N_1$ of polymers  with $L\ge 2$. 
Thus 
\begin{equation}
N_{m} = <N_{L=1}>+\sum_{L=2}^{L_{max}}L(exp(\alpha L+C_2)-exp(\beta\Delta L))e^{-\beta\Delta} e^{\beta\mu}
\label{eq:NmN1breakout} 
\end{equation} 
For $T>T_{c,2}$ , $\alpha$ is $>0$ and the sum increases exponentially with 
whereas if $T<T_{c,2}$ the sum converges even if $L_{max}\rightarrow \infty$ 
The second term on the right is largest when $\mu=0$ because the chemical 
potential of this
system must be negative to assure positive numbers $<N_L>$ at 
low temperatures. We denote the sum
\begin{equation}
{\cal{S}}(\beta) = \sum_{L=2}^{L_{max}}L(exp(\alpha L+C_2)-exp(\beta \Delta L))
\label{eq:}
\end{equation}
For use in analysis of experiments, we rewrite this in terms of densities,
introducing a factor in the entropy which accounts for the entropy of mixing
of the polymers in a volume $V$ as we did and describe in detail in references
\cite{quench,life2}giving 
\begin{equation}
N_{m}(v_p/V) = (v_p/V)<N_{L=1}>+{\cal{S}}'(\beta)e^{\beta\mu}
\label{eq:densityeq}
\end{equation}
where each term in the sum ${\cal{S}}'(\beta)$ has an added factor $L^{-3\nu}$ 
and $v_p=4\pi r_0^3/3$ is a microscopic volume describing the volume of the 
a polymer coil of length $1$. 
Rearranging 
we have
\begin{equation}
(1-<N_1>/N_{m})=(e^{-\Delta\beta}/{\rho v_p}) {\cal{S}}'(\beta)e^{\beta\mu}
\label{eq:predictlabel1}
\end{equation}
and 
\begin{equation}
{\cal{S}}'(\beta)= \sum_{L=2}^{L_max}L^{1-3\nu}(exp(\alpha L+C_2)-exp(\beta \Delta L))
\label{eq:sprime}
\end{equation}
${{S}}(\beta)'$ can be evaluated using the relations above for each reported 
temperature and polymer density reported for the observations and 
the left hand side of equation (\ref{eq:predictlabel1}) can be easily deduced for each sample from
the data as described in the next section.  $\rho$ is the total amino acid number
density which is reported in the data (different for each sample). In the case that
, at experimental temperatures,
${\cal{S}}'(\beta )(exp(-\beta\Delta)\rho v_p)$ is larger than 1  
, then a finite negative value
of $\mu$ can be computed from  (\ref{eq:densityeq}) and $<N_1>$ is predicted
to be zero.  On the other hand if  
${\cal{S}}'(\beta )/{\rho v_p}$ 
is smaller than 1, then 
we require that $\mu=0$ and the prediction is
\begin{equation}
(1-<N_1>/N_{m})=((exp(-\beta\Delta)/(\rho v_p)) {\cal{S}}'
\label{eq:prediction}
\end{equation}

\section{Application to analysis of the Mariana Trough  data}

We consider data from six samples from the campaign described in 
\cite{trench}. We  
used  values of $\{p_t\}$,  $<N_1>/N_{m}$ the total amino acid density
and the temperature of the emerging fluid in the analysis.
Numbers were reported for 11 amino acids so that $b=11$, (although of those
11, argenine was not detected in the samples from the smokers.  Argenine appeared in an ocean water sample
taken at the same depth.)
For each sample  
we estimate the
values of the apriori probabilities $\{p_t\}$ of the model by the relation
\begin{equation}
p_t = \sum_L <n_{L,t}>_{expt}/\sum_{L,t} <n_{L,t}>_{expt}
\label{eq:}
\end{equation}
which are easily determined for each observed smoker from Table 3 of
 \cite{trench}  as shown here in Table 1.  Values of $C_1, C_2, T_{c,2}$ 
the reported temperature of the effluent
are shown here in Table 2.  

\begin{table*}
\caption{Concentrations $\{p_t\}$ extracted from the six samples reported
in \cite{trench}}
\begin{tabular}{|c|c|c|c|c|c|c|c|c|c|c|c|}
\hline
Sample name & Asp& Thr &Ser &Glu &Gly &Ala &Val &Phe &Lys & His & Arg  \\ \hline \hline
1.D1435 CW & 0.11   &    .07  &0.33    &   .04  &0.23  &   0.13    &   .05  & 0.00    &   0.00    &   .03  & 0.00 \\ \hline    
2. D1435 CW  & .07  & 0.00     &  0.00   &   0.17   &   0.31    &   0.00    &   0.00   &    0.00    &  0.12   &   0.32   &    0.00  \\ \hline 
3.D1436 CW  & .09  & .06  &0.23   &   0.12    &  0.29   &    .09  & .05  & 0.00     &  .03  & .05  & 0.00   \\ \hline 
4. D1438 CW  &0.13   &    .09  &0.31     &  .09 & 0.39   &   0.12   &    0.00   &   0.00   &    0.00   &    .03  & 0.0   \\ \hline
5. D1440 CW  &0.19    &   0.00    &  0.25    &   0.00    &  0.54   &   0.00   &    0.00    &   0.00   &    0.00  &    0.00    &   0.00   \\ \hline 
 6. D1441 CW  & .10 & .06  &0.33     &  .03  &0.27  &    0.16   &    0.00   &    0.00   &    .02  & .022  & 0.00 \\ \hline   
\end{tabular}
\end{table*}

\begin{table*}
\caption{Values of $C_1,C_2,T_{c,2}$ extracted from the values in Table 1
, the fit described and illustrated in Fig. 1 and the reported temperatures
of the effluent from the smokers from which the 6 samples were taken as
reported in \cite{trench}.}
\begin{tabular}{|c|c|c|c|c|}
\hline
Sample  & $C_1$ & $C_2$ & $T_{c,2}$& $T_{observed}$   \\ \hline \hline
1. & 1.30799055   &  -2.61798072   &846.    &  710.67     \\ \hline                
2.  &  1.18290043 & -2.27813458     &  935.9   &   461.06       \\ \hline 
3.  & 1.3648520  & -2.7470862  &811.13   &   639.77       \\ \hline 
4.   &1.12157631   & -2.1479790 & 987.07  &803.7         \\ \hline
5.   &0.903600395    &  -1.64104235    &  1225.19    &  481.73   \\ \hline
6.  & 1.25609303 & -2.49383354  &881.37     &   802.25    \\ \hline
\end{tabular}
\end{table*}

In each of
the samples the temperatures were lower than $T_{c,2}$, the calculated
values of $(1/{\rho v_p}) {\cal{S}}'(\beta)$ are less than 1 and 
we take $\mu =0$ in equation (\ref{eq:predictlabel1}).  
 Some further details concerning the data extraction are described
in Appendix III.

\begin{figure}
\includegraphics[width=2.5in,angle=-90]{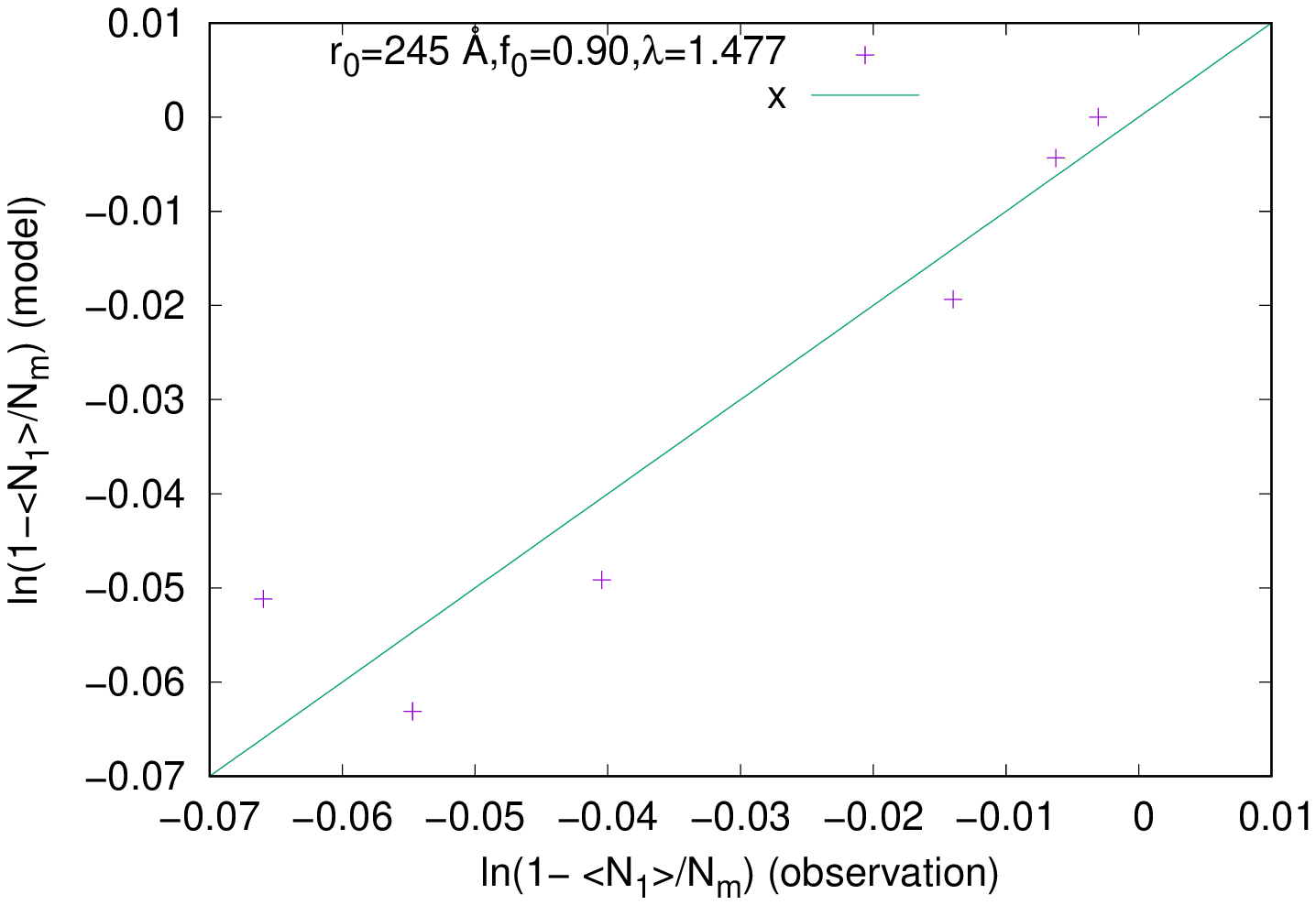}
\caption{\label{fig:BenFigp1}
Comparison of values of the left hand side of equation (\ref{eq:model}) with 
the calculated model values using the expression on the right hand side.
}
\end{figure}

It turns out that there is  discrepancy of several orders of magnitude
between equation (\ref{eq:prediction}) and the observations if $r_0$ is as
small as 2 angstroms (as reported for modern proteins in \cite{doniach}). 
We find agreement in order of magnitude with the observations if $r_0$ is
around 120 Angstroms.   
However the predictions span a larger sample-to-sample range than the observations
We account for that by assuming that, in addition,  there is a substantial
background fraction $f_0$ of bonded polymers that does not vary from sample to sample.
In summary we compare the predicted values:
\begin{equation}
\ln(1-<N_1>/N_{m})=\ln[((exp(-\beta\Delta)/(\rho v_p))(1-f_0){\cal{S}}'(\beta)+f_0]
\label{eq:model}
\end{equation}
With the observed values of the quantity on the left hand side of 
equation (\ref{eq:model}) on the righthand side we scaled the 
reported observations of the temperatures by a common factor $\lambda$
which was adjusted in the fitting of the righthand side to the
observed values.   In summary, we fit the observed values for the 
quantity on the left of (\ref{eq:model}) to the right hand side using
$r_0, f_0  $ and $\lambda$ as fitting parameters with results shown
in Figure \ref{fig:BenFigp1}. 
A further analysis of the very substantial uncertainties
associated with this comparison appears in Appendix III.

\section{Discussion and Conclusions}

At the temperature of ocean water at the depth of the Mariana Trough the equilibrium
distribution of polypeptide lengths would be nearly all monomers.  The 
concentrations reported in \cite{trench} show that, by contrast only about
2\%  of the amino acids emitted from the smokers are in monomeric form.  This could
be  because the observed amino acids have a biogenic origin, it could be
because the high temperature in the water emitted from the smokers before it is 
quenched has induced polymerization of the monomeric amino acids found in the minerals
near the mantle or it could be some combination of the two effects.  The second mechanism
is the one suggested as possibly initiating candidate polypeptides for starting
lifelike chemistry in the scenario in which a smoker was the site of the origin of life.
Our detailed analysis in this paper formulated predictions for the fraction of monomers
in the samples taken as reported in \cite{trench}.  The predictions resulting from
the assumption that all the observed polypeptide distribution arose from the second mechanism
gave  a larger sample to sample variation  than the one observed.  However if
we modelled the third possibility by assuming a, possibly biogenic, contribution to the 
monomer to polymer ratio which was sample independent, as in the third possibility above,
then we found rough agreement between the observations and the model. The fraction of
the material in the smokers which was independent of sample could  possibly be attributed
to mixture of ocean water with the water in the emissions.  It turned out that that 
fraction $f_0$needed to be  around 90\% in order to account for the data.
The other adjustment in the prediction which was required was an increase in the
factor $\exp(-\Delta/k_BT)/(\rho v_p))$ in equation (\ref{eq:prediction}). 

As discussed in Appendix III, we allowed for 3 effects which might account
for the remaining discrepancies between the observations and the model,
namely uncertainties in $\Delta$, in the temperatures of the hot fluids
emerging from the smokers and in $v_p$ parametrized by $\delta\Delta_{avg}$,
$\lambda$, and $r_0$ respectively.  We estimated $\delta\Delta_{avg}$
from data in \cite{martin1998free} and varied $\lambda$,$f_0$ and $r_0$ to
fit the model to the observations.  Details are in Appendix III. It turns
out that $f_0$ seems well established but that $r_0$ and the temperatures
are subject to large uncertainties as a result of this analysis.
The data used for the distribution of $\Delta$ values in this analysis
are limited to the report of peptide bond strengths in \cite{martin1998free}
for glycine-glycine bonds in pure water. 
On the other hand,  the observations were done in sea water, and we found no
data like that in \cite{martin1998free} for polypeptide bonds between 
the 11 other amino acids reported to be present.   
The factor $\Delta$ thus deserves, in our opinion, further attention as a 
possible
source of discrepancies in the comparison of the model with the observations.
We have not considered pH here. There is data available on the pH dependence
of the rate of peptide bond hydrolysis\cite{PHexpt} but we are here interested in the
free energy of the equilibrium bond and found only \cite{martin1998free}
which seems carefully done but extremely limited in scope. The big changes
in hydrolysis rates observed with pH might suggest similar dependence of
the equilibrium bond free energies which could possibly also help to account
for the discrepancies we report here.

We conclude that these comparisons of the predictions of our model  with the oceanographic data of reference 
\cite{trench} do not exclude the possibility that polypeptides produced in smokers
and then quenched could have produced the kind of repeated sampling and fixing
of polypeptides needed in the natural search for a combination leading to the 
initiation of life. The sharp contrast between the small fraction of 
amino acid monomers in the emitted fluids compared with the expected equilibrium distribution of nearly all monomers at the cold temperatures near the 
ocean floor itself
indicates that the polymer length distribution of those polypeptides is very 
far from equilibrium in the final state of the emitted fluids, as required
for any attribution of the origin of life to those fluids.

\section{Acknowledgements}
Aaron Wynveen is thanked for his careful reading of the manuscript and his
several useful suggestions. Professor Tomohiro Toki, coauthor of \cite{trench} 
is
thanked for his guidance in the interpretation of the data reported in that
paper. Early work on this project was supported in part by the United States National Aeronautics
and Space Administration (NASA) through grant NNX14AQ05G.

\section{Appendix I: Expression for $<N_L>$}

We imagine regrouping
polymers labelled by  $i$ and of length $L$ in order of their $s_L$ values
along a line of possible $s=s_L/L$ values from $0$ (degeneracy 1) to a maximum
of $\ln b$. There will be a number $N_{L,s}$ of polymers with the same value
of $s_L$ such that $\sum_{s=0}^{s=\ln b} N_{L,s}=N_L$ where $N_{L,s}$
is the number of polymers of length $L$ with $s$ value s.  The total number $W$
of microstates
of the set of polymers of length L is then
\begin{equation}
W=\prod_s {{(N_{L,s}+G_{L,s}-1!}\over{N_{L,s}!(G_{L,s}-1)!}}
\label{eq:}
\end{equation}
with
$G_{L,s}=e^{sL}$.

Take the $ln W$ giving a sum
\begin{equation}
\ln W = \sum_s \ln {{(N_{L,s}+G_{L,s}-1!}\over{N_{L,s}!(G_{L,s}-1)!}}
\label{eq:}
\end{equation}

The sum on $s$ will be different for each polymer $i$ among the
$i=1,...,N_L$ polymers of length $L$.  We take this into account as follows:
Divide by $N_L$ and sum on $i$:
$$
\sum_s ...
 = \int ds (1/N_L)\sum_i \delta (s+\sum_t (n_{L,i,t}/L)\ln(n_{L,i,t}/L))...
$$
\begin{equation}
= \int ds{\cal{D}}(s)...
\label{eq:}
\end{equation}
defining a 'density of states' ${\cal{D}}_L(s)$
\begin{equation}
{\cal{D}}_L(s)=(1/N_L) \sum_i \delta (s+\sum_t (n_{L,i,t}/L)\ln(n_{L,i,t}/L))
\label{eq:}
\end{equation}
which is the density of s per polymer in the set of $N_L$ polymers of
length $L$.  The density could be evaluated in a simulation by constructing
a histogram.  However we  get the average over the
probability distribution of the $\{n_{L,it}\}$,
using the distribution given in (\ref{eq:PL})
\begin{equation}
{\cal{P}}_{L}={{L!\prod_{t=1}^b p_t^{n_{L,i,t}}}\over{\prod_{t=1}^b n_{L,i,t}!}}
\label{eq:}
\end{equation}
for each polymer $i$. Within each summand in ${\cal{D}}_L(s)$
the average will be the same so that, denoting the average by $<...>$
we have

$$<{\cal{D}}_L(s)>=
$$
\begin{equation}
\sum_{\{n_{L,i,t}\}\ni \sum_t n_{L,i,t}=L}
\delta (s+\sum_t (n_{L,i,t}/L)\ln(n_{L,i,t}/L))
\label{eq:}
\end{equation}
$$
\times {\cal{P}}_{L}(\{n_{L,i,t}\})
$$.

The $N_L$ 's cancel out and this density of states is independent
of $N_L$.
Thus

\begin{equation}
<\ln W> = \int ds <{\cal{D}}_L(s)>\ln \left [{{(N_{L,s}+(e^{Ls}-1))!}\over{N_{L,s}!(e^{sL}-1)!}}\right]
\label{eq:}
\end{equation}
Now one can maximize the integrand with respect to $N_{L,s}$ inside the
integral on s
using Stirling's approximation because, within an $s$ interval $ds$ the
degeneracies are all the same.  This cannot be done when $L=1$ and
will not be accurate unless $N_{L,s}$  is substantially larger than $1$
when $L>1$.
Then, inside
the integral on $ds$ we have
\begin{equation}
\overline{N_{L,s}} = {{e^{sL}-1}\over {e^{-\mu_\beta -\beta\Delta(L-1)}-1}}
\label{eq:}
\end{equation}
Multiply by  $<{\cal{D}}(s)$  and integrate over $s$ to get the
expected value of $N_L$ denoted $<\overline{N_L}>$. The only $s$ dependence
in $\overline{N_{L,s}}$ is in the factor $e^{sL}$ giving
\begin{equation}
<\overline{N_L}>={{<e^{sL}>-1}\over {(e^{\beta\Delta(L-1)-\mu\beta}-1)}}
\label{eq:avNL}
\end{equation}
for $L\ge 2$ and in which
\begin{equation}
<e^{sL}>=\sum_{\{n_{L,t}\}\ni \sum_t n_{L,t}=L}exp\left[\ln(L-1!)-\sum_t ln (n_{l,t}!)\right]{\cal{ P}}_L
\label{eq:etosL}
\end{equation}
as cited in the main text equations (\ref{eq:avNLb}) and (\ref{eq:etosLb}).
\section{Appendix II. Expression for $N_1/N_{m}$}

As in \cite{life2} we
introduce a parameter
\begin{equation}
\alpha = \beta\Delta(1-T/T_{c,2})
\label{eq:alpha}
\end{equation}
 which is formally the same as in \cite{life2}
but with a different definition of $T_{c,2}$
(The overall sign on the right hand side of equation (28)of
\cite{life2} is in error and should be +.). Values of $T_{c,2}$
are different for the six samples studied and are listed in Table 1.

Within the model, there is no energy
associated with totally unbonded monomers so the method of maximization
of the entropy does not provide a prescription for its value.
We regard the $\mu$ in the previous equations as determined from
the number, termed $N_{m}-N_1$ of polymers in with $L\ge 2$.
Thus
\begin{equation}
N_{m}=<N_{L=1}>+\sum_{L=2}^{L_{max}}L(exp(\alpha L+C_2)-e^{(\beta\Delta)})e^{-\beta\Delta} e^{\beta\mu}
\label{eq:Nem}
\end{equation}
For $T > T_{c,2}$ , $\alpha$ is $> 0$ and the sum increases exponentially with
$L_{max}$
whereas if $T < T_{c,2}$ the sum converges even if $L_{max}\rightarrow \infty$
The second term on the right is largest when $\mu=0$ because the chemical
potential of this
system must be negative to assure positive numbers $N_L > 0$ at
low temperatures. We denote the sum
\begin{equation}
{\cal{S}}(\beta) = \sum_{L=2}^{L_{max}}L(exp(\alpha L+C_2)-exp(\beta \Delta L))
\label{eq:}
\end{equation}
For use in analysis of experiments, we rewrite this in terms of densities,
introducing a factor in the entropy which accounts for the entropy of mixing
\cite{quench,life2}.
\begin{equation}
N_{m}(v_p/V) = (v_p/V)<N_{L=1}>+{\cal{S}}'(\beta)e^{\beta\mu}
\end{equation}
where each term in the sum ${\cal{S}}'(\beta)$ has an added factor $L^{-3\nu}$
and $v_p=(4\pi r_0^3/3$ is a microscopic volume describing the effective 
volume occupied by one monomer. 
Rearranging
we have
\begin{equation}
(1-<N_1>/N_{m})=(1/{\rho v_p}) {\cal{S}}'(\beta)e^{\beta\mu}
\label{eq:predictwithmu}
\end{equation}
and
\begin{equation}
{\cal{S}}'(\beta)= \sum_{L=2}^{L_{max}}L^{1-3\nu}(exp(\alpha L+C_2)-exp(\beta \Delta L))
\label{eq:sprimea}
\end{equation}
${\cal{S}}(\beta)'$ can be evaluated using the relations above for each reported
temperature and polymer density reported for the observations and
the left hand side of equation (\ref{eq:predictwithmu}) can be easily
deduced for each sample from
the data as described in the section \label{sec:data}.  $\rho$ is the total amino acid number
density which is reported in the data (different for each sample). In the case that
, at experimental temperatures,
${\cal{S}}'(\beta )/{\rho v_p}$ is larger than 1 
,  a finite negative value
of $\mu$ can be computed from this equation and $<N_1>$ is predicted
to be zero.  Under the conditions in the six smoker samples reported
in \cite{trench} a macroscopic number of unbonded monomers is reported
 so that ${\cal{S}}'(\beta )(e^{-\beta\Delta}{\rho v_p}$
must be smaller than 1, and
the prediction is
\begin{equation}
(1-<N_1>/N_{m})=(e^{-\beta\Delta}/{\rho v_p}) {\cal{S}}'(\beta)
\label{eq:predictiona}
\end{equation}
with ${\cal{S}}'(\beta)$ defined in (\ref{eq:sprimea})
(\ref{eq:predictiona}) appears in the main text as
(\ref{eq:predictlabel1}) and (\ref{eq:sprimea}) as 
(\ref{eq:sprime}).
 
\section{Appendix III: Discussion of errors and uncertainties in predictions}

We only have data from six samples and have been warned (by an author of 
\cite{trench}) that the
reported temperatures of the fluids leaving the smokers are quite uncertain
and probably too low.  Further there is reported to be a significant 
dilution of the samples by ambient ocean water.  In addition there 
are very few laboratory reports of the peptide bond free energy $\Delta$
in pure water and that parameter plays an important role in the model.
As explained in the main text, we were forced to assume a large component
of the observed values of $(1.-<N_1>/N_{m})$ to a 'background' common
to all the samples and characterized by the parameter $f_0$. That 
'background' could be due to dilution or to a large biogenic component
in the emitted fluids from the smokers.

In attempting to quantify the resultant uncertainties in the results we
begin with the report \cite{martin1998free} on  experimental values of the 
parameter $\Delta$ for glycine-glycine bonds in six different peptide bond
contexts.  The values reported are in order of magnitude agreement 
with one another but vary significantly.  To obtain the data exhibited
in  Figure \ref{fig:BenFigp1} 
we used the average $\Delta_{avg}$ of the values reported in Table 1 
of \cite{martin1998free}
in evaluating the right hand side of (\ref{eq:model}) and then
minimized the sum of mean squared differences between the experimental left and
modelled right hand sides of (\ref{eq:model}) with respect to $r_0$,
$f_0$ and the scale factor $\lambda$ for the temperatures. 
To get an idea of the uncertainties associated
with that procedure we then repeated the procedure, starting with values
of $\Delta$ which were $\Delta_{avg}\pm \delta\Delta$ where 
$\delta\Delta$ was the standard deviation of the values reported in 
\cite{martin1998free} from their average. For those other two values of $\Delta$
we repeated the minimization of the mean square deviation of the
model values in equation (\ref{eq:model}) from the observed
data for all six samples with respect to $R_0$, $f_0$ and
$\lambda$ for the temperature. The result enabled an estimate of the 
uncertainties in our
determinaton of the parameters $R_0$, $f_0$ and
$\lambda$. Results are shown in Table III.  They indicate that $f_0$ may be 
quite stably fixed but that the values of $r_0$ and the temperatures
are subject to very large uncertainties.

\begin{table}[b]
\caption{Ranges of the values of the parameters $r_0,f_0$ and $\lambda$
determined as described in this Appendix.}
\begin{tabular}{|c|}
\hline
$ 106 \AA \leq r_0  \leq 245.\AA$ \\ \hline
.90 $\leq f_0 \leq $0.94  \\ \hline
.48 $\leq \lambda \leq$ 1.48 \\ \hline
\end{tabular}
\end{table}

The procedure is only taking account of the known dependence of the 
input parameter $\Delta$ from the measurements of glycine-glycine peptide
bonds on various chemical contexts.  $\Delta$ is also expected to be
different for different pairs of amino acids among the 11 amino acids
detected in the smokers and that is expected to increase the deduced 
uncertainties. There could be further variance arising from differences in
pH and saline content in smoker fluids as compared with the pure water of
the experiments reported in \cite{martin1998free}. Comprehensive
laboratory data on peptide bond energies in contexts relevant
to the smoker environment would be welcome.

\end{document}